# Flat Helical Nanosieves

*Shengtao Mei[1,3][†], Muhammad Qasim Mehmood[1,5][†], Sajid Hussain[1][†], Kun Huang[2], Xiaohui Ling[1,4], Shawn Yohanes Siew[1], Hong Liu[2], Jinghua Teng[2], Aaron Danner[1] and Cheng-Wei Qiu[1,3][*]*

[1]Department of Electrical and Computer Engineering, National University of Singapore, 4 Engineering Drive 3, Singapore 117583, Singapore

[2]Institute of Materials Research and Engineering Agency for Science Technology and Research (A*STAR), #08-03, 2 Fusionopolis Way, Innovis, Singapore 138634

[3]Graduate School for Integrative Sciences and Engineering, National University of Singapore, Centre for Life Sciences (CeLS), #05-01, 28 Medical Drive Singapore 117456, Singapore

[4]College of Physics and Electronic Engineering, Hengyang Normal University, Hengyang 421002, China

[5]Information Technology University (ITU), 346-B, Ferozepur Rd, Lahore 54600, Pakistan.

[†]These authors contributed equally to this work

[*]Correspondence and requests for materials should be addressed to C.W. Q. (email: chengwei.qiu@nus.edu.sg).

**ABSTRACT:**

**Compact and miniaturized devices with flexible functionalities are always highly demanded in optical integrated systems. Plasmonic nanosieve has been successfully harnessed as an ultrathin flat platform for complex manipulation of light, including holography, vortex generation and non-linear processes. Compared with most of reported single-functional devices, multi-functional nanosieves might find more complex and novel applications across nano-photonics, optics and nanotechnology. Here, we experimentally demonstrate a promising roadmap for nanosieve-based helical devices, which achieves full manipulations of optical vortices, including its generation, hybridization, spatial multiplexing, focusing and non-diffraction propagation etc., by controlling the geometric phase of spin light via over 121 thousands of spatially-rotated nano-sieves. Thanks to such spin-conversion nanosieve helical elements, it is no longer necessary to employ the conventional two-beam interferometric measurement to characterize optical vortices, while the interference can be realized natively without changing any parts of the current setup. The proposed strategy makes the far-field manipulations of optical orbital angular momentum within an ultrathin interface viable and bridges singular optics and integrated optics. In addition, it enables more unique extensibility and flexibility in versatile optical elements than traditional phase-accumulated helical optical devices.**

# 1. Introduction

Benefitting from the replacement of phase accumulation in bulk materials with the abrupt interfacial phase discontinuity, metasurface is considered as a promising 2D metamaterial to design integrated optical devices by controlling electromagnetic wave's phase, amplitude and polarization state in a desired manner[1-5]. Varieties of metasurface devices have been fabricated to demonstrate their powerful ability based on the generalized laws of reflection and refraction. Examples include optical vortex plate [6], ultrathin flat lens [7-10], propagating wave to surface wave convertor [11, 12], broadband quarter-wave plate [13], directional surface plasmon polaritons excitation [14], and meta-hologram[15-20]. Especially for the Pancharatnam-Berry metasurface [21], it transfers the interfacial phase distribution into orientations of base units, which largely simplifies the design process of optical antenna compared with those metasurfaces based on spatially varying geometric parameters (e.g. V-shaped metasurface, slot antenna array metasurface [22-24]). Thus, metasurface provides new opportunities and strategies to develop ultrathin optical devices that are compatible to integrated and compact systems. However, most of the aforementioned metasurface based optical devices are limited to single-specific-purpose devices, which restricts their application value.

On the other hand, traditional integrated optical devices, which are always fabricated by direct laser writing on bulky optical material, accumulate phase change through light's propagation in the refractive optical material. Therefore, smooth phase change requires continuous thickness change, which demands stringent spatial resolution. The smaller the device gets, the more challenging requirements on the spatial resolution will be. It would be more difficult for traditional strategy when complex surface is required. Specifically, the topological charge dependent thickness of the Spiral Phase Plate will induce more uncertainties on the required performance and also increase the total thickness of the whole device[25-27], which may not be compatible with integrated optical systems.

Recently, methods of microscopic generation of optical vortex beam [28], including plasmonics [29], integrated optics [30, 31], spiral zone plate [32], optical coordinate transformations [33], and fiber optics [34], have been developed. Potential strategies of new generation method are proposed [35]. However, more complex manipulations of the generated optical vortex beam are always required in practical applications such as optical tweezers/optical spanner [36-38], optical communication [39-42], and even Stimulated Emission Depletion (STED) Microscopy [43]. For example, optical spanners that are used to rotate small particles are realized by highly focused vortex beams [36, 37]. Up to now, most of the metasurfaces generated vortex beams are Laguerre-Gauss-like beams, which will diverge after its generation [6, 44]. Nondiffracting vortex beams are desired, e.g. trapping of long and thin objects such as rods and E-coli, simultaneously trapping of both high and low refractive index particles [45-48], and implementing the Bessel-beam STED (BB-STED) [43], while additional manipulations are difficult to implement after such micro-scale generation. Here, we would like to achieve complete control of optical vortex beam's generation, nondiffraction, focusing and characterization via multifunctional helical nanosieves, which may bridge the fields of singular optics and integrated optics.

In this paper, a series of complex controls of optical vortex beams are reported based on flat helical devices. Without loss of generality, we employ the Babinet Pancharatnam-Berry (PB) gold metasurface, or called nanosieve array, as the platform to achieve varieties of multiple orbital angular momentum (OAM) loaded light control. Nanosieve for dealing spin photons and circularly polarized (CP) lights are particularly demanded with higher signal-to-noise ratio (SNR) (SNR = 10 log$_{10}$ ($P_d$/$P_{ud}$) dB) by increasing desired transmission power ($P_d$: cross-CP transmitted power, $P_{ud}$: co-CP transmitted power), which cannot be addressed by reflection-type ones (though with higher efficiency). Strategy of improving efficiency has been investigated in many reported works, our work, however, provides a comprehensive recipe for designing spin-dependent multifunctional helical nanosieves enabling complete control of vortex manipulations based on our available fabrication conditions verified with fairly good experimental results. Unlike the traditional nanofabricated OAM optical elements, our multifunctional flat helical nanosieves can not only manipulate optical vortex, but also show unique extensibility and flexibility during the process of fabrication and applications.

## 2. Materials and Methods

The two processes of superimposing phase profiles for processing OAM are depicted in Figure 1(a). The first row presents the process of superimposing phase profiles of two concentric axicons and two concentric spiral phase plates. The second row presents the process of superimposing phase profiles of two concentric lenses and two concentric spiral phase plates. The two concentric regions in Figure 1(a) correspond to the Region I and II in Figure 1(b), respectively. The insets in Figure 1(a) show the SEM images of the same position on the fabricated samples of the two flat helical devices in which the nano-void's different orientation angles are resulted from the different phase value of helical axicon and vortex lens. The flat helical nanosieve is fabricated by patterning on gold film with perforated nano-voids: Babinet-inverted (complementary) photon nanosieves on quartz substrate. It is composed of nanosieves with spatially varying orientation in two different regions marked by I and II [cf. Figure 1(b)]. The radii of Region I and II are kept as 0.5~45 μm and 50~100 μm, respectively. In our design, the nanosieve has its length $l = 150$ nm, width $w = 75$ nm, and depth 60 nm. All of the nanosieves are equally spaced ($d_1 = 500$ nm) on the concentric circles, and all the circles are separated with the same period ($d_2 = 500$ nm) [cf. Figure 1(b) inset]. A scanning electron microscopy (SEM) image shown in Figure 1(c) and its insets in Figure 1(a) clearly show the rotating rectangular nanosieves. There are 121,486 nanosieves in total with different orientation angles. The abrupt phase delay of the flat helical device is realized by controlling the orientation angle $\varphi$ of the nanosieves relative to the $x$ axis. As a result, the designed phase distribution $\Phi\,(x,y) = \pm 2\varphi(x,y)$ ("+": right circularly polarized (RCP) incident light; "-": left circularly polarized (LCP) incident light) can be controlled precisely and smoothly on nanosieves. The basic idea is that the two regions can process the loaded optical vortex independently by integrating different manipulating phases. Figure 1(d) depicts the illumination of the two kinds of proposed flat helical devices: Helical Axicon and Vortex lens, which shows the basic functions (hybridizing and multiplexing) of these helical nanosieves. The experimental setup for characterization is shown in Figure 6S. The fabrication process can be found in the Experimental Section.

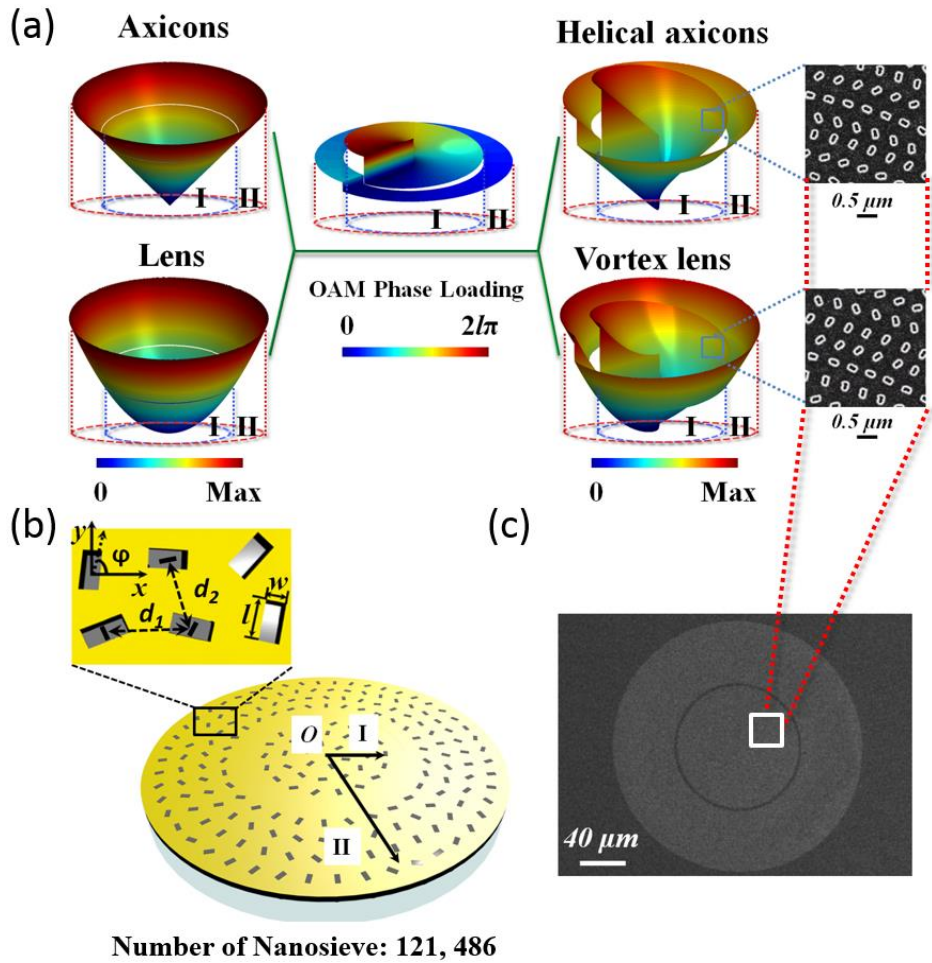

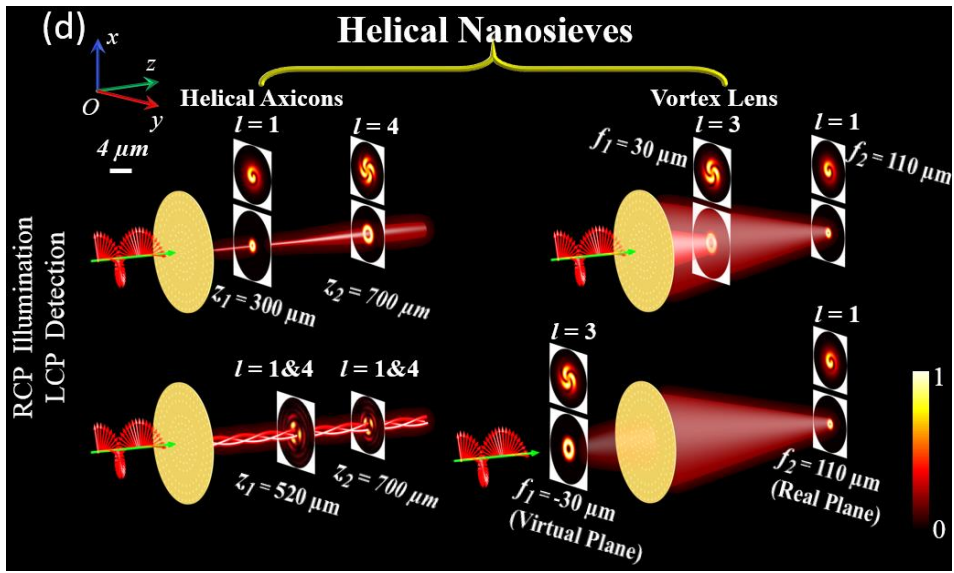

**Figure. 1.** Procedure of superimposing phase profiles, schematic diagram and SEM images of the two-region flat helical nanosieves, and functioning demo of the vortex nanosieves for processing OAM. (a) Processes of obtaining the phase profiles of helical axicon and vortex lens by combining spiral phase with different manipulation phases. Insets: SEM images of the same position on the two fabricated vortex nanosieves which correspond to different designs. (b) Schematic diagram of the two-region helical nanosieves consisting of rotating perforated photon nano-void---complementary nano-

antennas (shown in the inset) --- on the 60 nm gold film. The device contains two regions (Region I and Region II) with a gap (5 μm) between them. Region I is 0.5~45 μm, and Region II is 50~100 μm. The value of the parameters in the inset are $d_1 = d_2 = 500$ nm, $l = 150$ nm, and $w = 75$ nm. φ is the local orientation angle of each nano-void with respect of *x* axis; (c) SEM image of one two-region helical device. All of the designs are fabricated in this shape. The orientation angle of every nano-void is determined by specific design. Two of them are shown in the insets of (a); (d) Illumination of different functional (hybridizing and multiplexing) flat helical devices (Helical Axicons and Vortex Lens) with simulated intensity profiles and the corresponding interference patterns.

## 3. Results and Discussion

### 3.1. Cascaded Bessel Beam Generator

We first show two-region helical axicon to generate a piecewise Bessel beam---cascaded nondiffraction optical vortex beams, which is obtained by superimposing the phase profiles of axicon and spiral phase plate. The axicon projects a point source on to a line segment along the optical axis, whose length is regarded as the depth of focus (DOF) [49]. Inspired by this, helical axicon, based on conventional micro fabrication, was designed to achieve the high order Bessel-Gauss-like beams[50]. For an axicon with opening angle $\beta = \tan^{-1}(\frac{R}{\text{DOF}})$ ($R$ is the radius of the axicon), its phase delay increases monotonically with the distance from the center. In other words, it is a conical radial phase distribution expressed as $\phi_A = \frac{2\pi}{\lambda} r \sin\beta$, where $\lambda$ is the working wavelength and $r$ represents the distance between the nano-sieve and the center of the device. While for a spiral phase plate with topological charge $l$, the phase profile is $\phi_S = l \tan^{-1}(\frac{y}{x})$. Hence, the overall phase profile of a helical axicon is $\Phi = \phi_A + \phi_S$, so the orientation angle distribution of the nano-sieves is $\varphi = \varphi_A + \varphi_S = \pm\frac{\Phi}{2} = \pm\frac{\phi_A + \phi_S}{2}$. Measured results of single-region helical axicons' field distributions along the propagation direction can be found in the supplementary materials. Meanwhile, their corresponding radially averaged intensity profiles of different cross-sections are well fitting with different $l^{\text{th}}$ order Bessel functions [cf. Figure 1S]. In the two-region designs, the orientation angle distribution of the nano-sieves can be rewritten in the following details:

$$\varphi(r) = \begin{cases} \pm\frac{\pi}{\lambda}r\sin\beta_1 \pm \frac{l_1}{2}\tan^{-1}\left(\frac{y}{x}\right) & 0.5 \leq r \leq 45\ \mu m \\ \pm\frac{\pi}{\lambda}r\sin\beta_2 \pm \frac{l_2}{2}\tan^{-1}\left(\frac{y}{x}\right) & 50 \leq r \leq 100\ \mu m \end{cases} \quad (1)$$

The above distribution can be considered as two concentric helical axicons with opening angle $\beta_1, \beta_2$ and topological charge $l_1, l_2$.

In a step-by-step process, we first investigate the case of the two regions with the same opening angle ($\beta_1 = \beta_2 = 6°$) and different topological charge ($l_1 \neq l_2$). Two concentric sub-helical axicons will generate two cascaded non-diffracting fields (one hybridizing Bessel beam) but with different radius because of the different helical phase gradient encoded onto them [cf. Figure 2]. The inner sub-helical axicon generates a non-diffracting field from 0 μm to 430 μm, while the outer one generates a non-diffracting field from 470 μm to 950 μm. The agreement concerning the propagating behavior of the light field emerging from the two flat helical axicons between the measured and simulated data is fairly good (refer Figure 2S for the simulation results). Figure 2(a) shows the mechanism of the two-region helical axicon, which clearly shows that the two regions can manipulate OAM independently. Figure 2(b) (c) show the experimental results of the two designed helical axicons. The measured intensity profiles and interference patterns are shown at the right side of Figure 2(b) (c). Both the two devices are designed for RCP illumination and LCP detection. Based on the spin-dependent phase response of the device, the output beams carry opposite chirality to that of the incident beam, because of which there is no restriction on the incident light beam's waist. However, when considering conventional micro-optical helical axicon [50] (combination of spiral phase plate and axicon), the incident Gaussian beam's waist is limited by the size of the device, which may impose restrictions on real applications because an extra focusing lens is needed to confine the incident beam. In characterization, circular polarizer is used at the transmission side to filter out the background light (light with the same chirality as that of the incident beam). Only light with opposite chirality is captured by Charge Coupled Device (CCD). When the background light is partially allowed to transmit by tuning the circular polarizer, a single-beam interferometric scheme allows for the unambiguous characterization of the phase structure of any encoded OAM state. The measured interference pattern results have shown good agreement with the simulation ones, which validates such an embedded approach as a simple and stable characterization technique. This expandable multi-region flat helical axicons can be really complex to fabricate by conventional techniques by direct laser writing using femtosecond laser nanopolymerization.

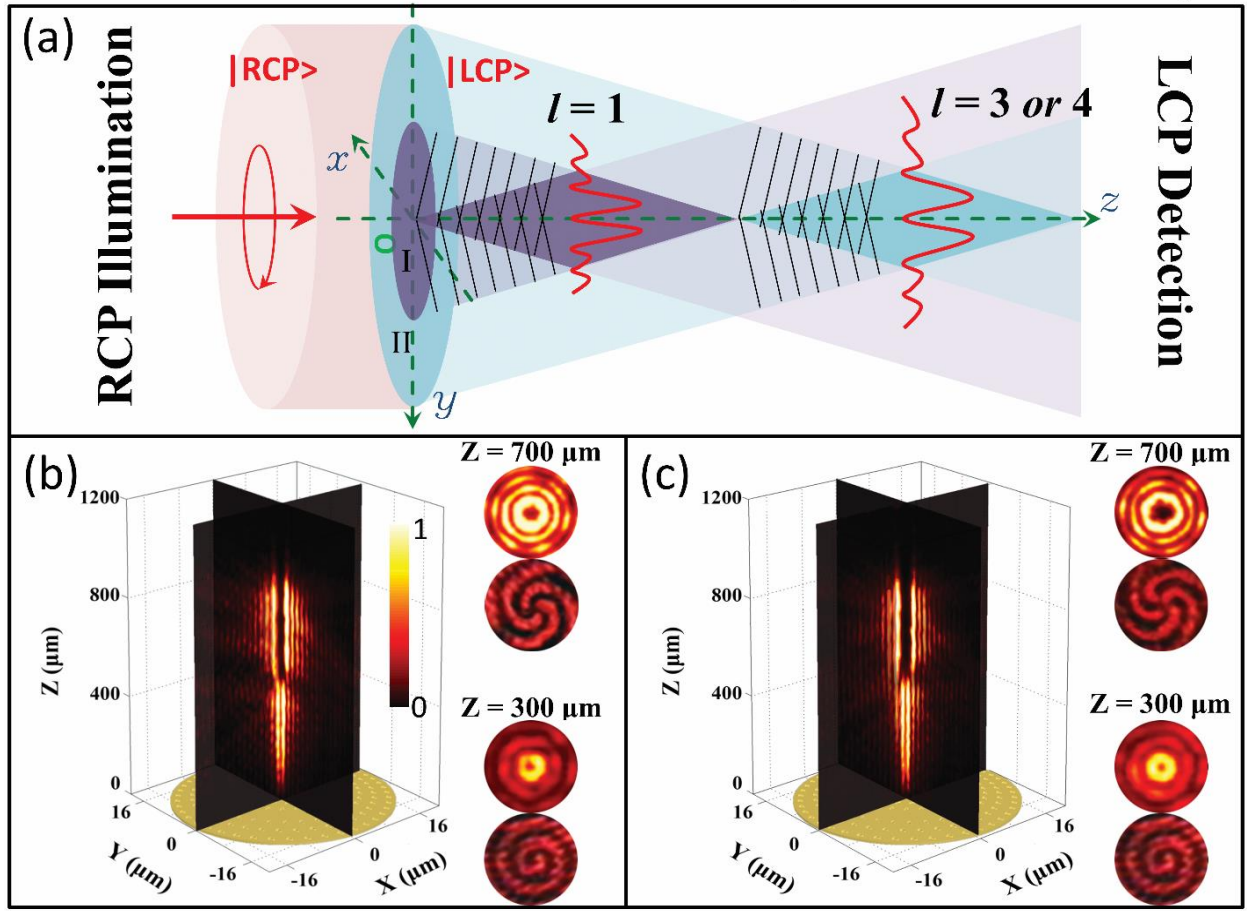

**Figure. 2.** Examples of two-region helical axicons to generate cascaded Bessel beams (hybridizing nondiffracting vortices). (a) Mechanism of the two-region helical axicons. (b) (c) Measured far-field distribution along the optical path for helical axicon containing two regions encoded with the same opening angle ( $\beta_1 = \beta_2 = 6°$) and different topological charge ( $l_1 \neq l_2$ ). (b) $l_1 = 1$, $l_2 = 3$. (c) $l_1 = 1$, $l_2 = 4$. The right-side images in (b) and (c) show the intensity profiles and interference patterns of the beam's x-y cross-sections within different nondiffracting field segment for each design at z = 300 μm and z = 700 μm.

**3.2. Helicon Wave Generator**

Two-region helical axicons with two different opening angles (helicon wave generators) are further investigated to show flexibility of this flat fabrication strategy [cf. Figure 3]. As validated in the single region cases [cf. Figure 1S], nondiffracting field generated by inner sub-helical axicon with $\beta_1 = 3°$ covers the nondiffracting field generated by the outer sub-helical axicon with $\beta_2 = 6°$. The range of the covered-field is from 470 μm to 950 μm. Two designs are presented in Figure 3, in which the two nondiffracting beams merge together to form a helicon wave [51] showing the rotating interference patterns during their propagation [cf. Figure 3S for the simulation results]. To be more specific, Figure 3(a) exhibits the experimentally detected helicon wave resulted from the superposition of the 1st order and the 3rd order Bessel beams. Their two-lobe intensity images rotate along the propagating direction as depicted in the transverse profiles [cf. right side of Figure 3(a)]. Similarly, the rotational three-lobe intensity images [cf. Figure 3(b)] correspond to the superposed patterns of the 1st order and the 4th order Bessel beams [52]. The total field can be decomposed into the contributions from the two sub-helical axicons, given respectively as:

$$U_{l_1}(r, \varphi, z) \propto J_{l_1}(k_{1r}r) e^{i(k_{1z}z + l_1\varphi)} \quad (2.a)$$

$$U_{l_2}(r, \varphi, z) \propto J_{l_2}(k_{2r}r) e^{i(k_{2z}z + l_2\varphi)} \quad (2.b)$$

where $k_{mr} = k_0 \sin \beta_m$, $k_{mz} = k_0 \cos \beta_m$ $\left(m = 1 \text{ or } 2, \text{ and } k_0 = \frac{2\pi}{\lambda}\right)$. If only considering the azimuthal ($\varphi$) and longitudinal ($z$) component, the superposed field is given by the following proportionality:

$$(U_{total})_{\varphi,z} = (U_{l_1} + U_{l_2})_{\varphi,z} \propto e^{i(k_{1z}z + k_{1z}z)}(1 + e^{i((k_{2z}-k_{1z})z + (l_2-l_1)\varphi)}), \quad (3)$$

from which we can obtain the corresponding interference pattern as the following proportionality:

$$I_{\varphi,z} \propto (1 + \cos((k_{2z} - k_{1z})z + (l_2 - l_1)\varphi)). \quad (4)$$

As a result, the multi-lobe pattern is a direct echo of the topological charge difference ($l_2 - l_1$) between the two high-order Bessel beams. The rotation of the interference pattern during propagation can be explained by its longitudinal dependence. The angular velocity of $I_{\varphi,z}$ along the propagation axis is fixed at a rate given by:

$$\frac{d\varphi}{dz} = -\frac{k_{2z}-k_{1z}}{l_2-l_1}. \tag{5}$$

So essentially, the longitudinal wavenumber difference ($k_{2z} - k_{1z}$) between the two Bessel beams, which is directly determined by the different opening angle $\beta$ of the two Bessel beams, leads to the behavior of longitudinal rotation. If two Bessel beams, having the same $k_z$, are somehow superposed together (not in our designs), the multi-lobe intensity will propagate without rotation ($\frac{d\varphi}{dz} = 0$), although the total OAM is not zero [53]. On the other hand, the $2l$-lobe pattern which is generated by superposing an $l^{th}$ order Bessel beam and its mirror image ($-l^{th}$ order Bessel beam) will rotate although the total OAM is zero ($\frac{d\varphi}{dz} \neq 0$). In addition, smaller interval between the two topological charges means larger angular velocity. However, the longitudinal period of the intensity pattern is $\Delta z = \frac{2\pi/(l_2-l_1)}{d\varphi/dz} = \frac{2\pi}{k_{1z}-k_{2z}}$ which is obviously independent of topological charge. So the two designs should have the same longitudinal period (~154 μm) theoretically, and it is also demonstrated by the experiment results [cf. Figure 3(c)]. The 1st row in Figure 3(c) shows the cross-section intensities along propagating direction from 600 μm to 760 μm for the design of $\beta_1 = 3°, l_1 = 1$, and $\beta_2 = 6°, l_2 = 3$, and the 2nd row shows the cross-section intensities along propagating direction from 600 μm to 760 μm for the design of $\beta_1 = 3°, l_1 = 1$, and $\beta_2 = 6°, l_2 = 4$. The two 3D schematic diagrams in Figure 3(c) depict the two helicon waves ($\Delta l = l_2 - l_1 = 2$ and 3) within one period. Both the designs clearly show one complete longitudinal period within 160 μm, and it confirms that smaller topological charge interval leads to larger angular velocity simultaneously. This nondiffracting and constant-rate rotating intensity distribution suggests that such waves may find applications requiring both precise alignment and distance/speed measurements [51].

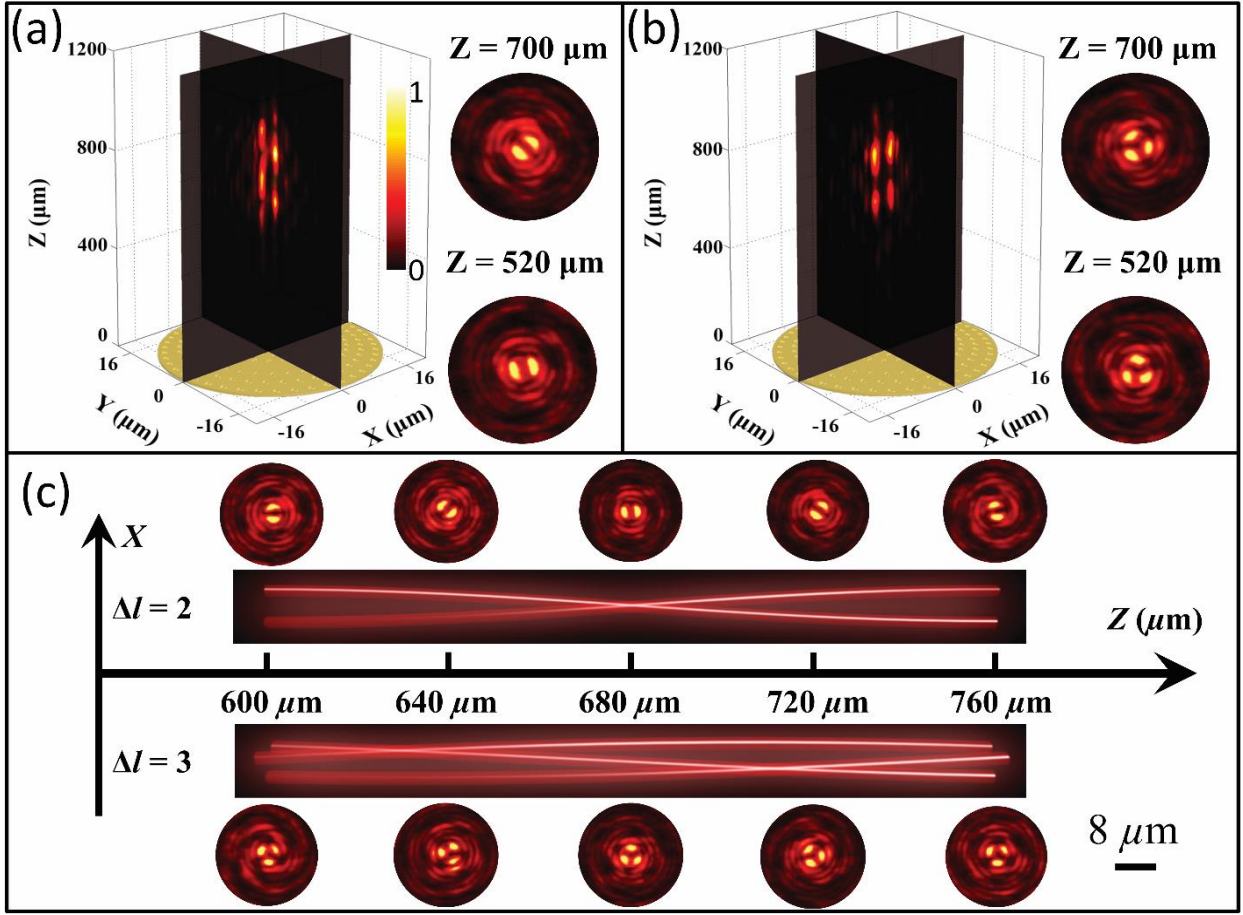

**Figure. 3.** Examples of helicon wave generators (spatially multiplexed nondiffracting vortices). Measured far-field longitudinal intensity distributions for two-region helical axicons encoded with different opening angles ($\beta_1 \neq \beta_2$) and different topological charges ($l_1 \neq l_2$). (a) $\beta_1 = 3°, l_1 = 1,$ and $\beta_2 = 6°, l_2 = 3$. (b) $\beta_1 = 3°, l_1 = 1,$ and $\beta_2 = 6°, l_2 = 4$. The right-side images in (a) and (b) show the intensity profiles of the beam's x-y cross-sections within the nondiffracting field for each design at z = 520 μm and z = 700 μm. (c) The rectangular images are two 3D schematics showing two helicon waves ($\Delta l = l_2 - l_1 = 2 \ and\ 3$) within one period (from 600 μm to 760 μm). The circular images represent the measured cross-section intensities at intervals of 40 μm along propagating direction (*z*) from 600 μm to 760 μm. The 1st row corresponds to the design of $\beta_1 = 3°, l_1 = 1,$ and $\beta_2 = 6°, l_2 = 3$. The 2nd row corresponds to the design of $\beta_1 = 3°, l_1 = 1,$ and $\beta_2 = 6°, l_2 = 4$. Both of the last intensity patterns recover the first ones which lead to the same longitudinal period of 160 μm.

Such helicon wave generators are fabricated by employing the same process as previous cases. The only difference lays in the phase distribution on the nanosieves. However, such phase distribution means far more complex surface polish processing in traditional laser writing technique. Meanwhile, it is worth pointing that by employing high-efficiency nanosieves one can easily have a hybrid Bessel beam merged from two beams with opposite chirality by shining linearly polarized light, as long as the two sub-helical axicons are designed for opposite circular polarizations. Then, demultiplexing the multiplexed beams can be easily implemented by a circular polarizer.

### 3.3. Multi-foci Vortex Lens

We also realized a multi-foci vortex lens to generate highly focused optical vortex beams, which is obtained by superposing the phase profiles of lens and spiral phase plate. The phase profile of a focusing lens is $\phi_L = \frac{2\pi}{\lambda}\left(\sqrt{r^2 + f^2} - f\right)$, where $f$ is the designed focal length for each sub-vortex lens. In all, the total phase profile of a vortex lens is $\Phi = \phi_L + \phi_S$, so the orientation angle distribution of the nano-voids is $\varphi = \varphi_L + \varphi_S = \pm\frac{\Phi}{2} = \pm\frac{\phi_L + \phi_S}{2}$. In the two-region design, the orientation angle distribution of the nano-voids can be rewritten in more detail as following:

$$\varphi(r) = \begin{cases} \pm\frac{\pi}{\lambda}\left(\sqrt{r^2 + f_1^2} - f_1\right) \pm \frac{l_1}{2}\tan^{-1}\left(\frac{y}{x}\right) & 0.5 \leq r \leq 45 \ \mu m \\ \pm\frac{\pi}{\lambda}\left(\sqrt{r^2 + f_2^2} - f_2\right) \pm \frac{l_2}{2}\tan^{-1}\left(\frac{y}{x}\right) & 50 \leq r \leq 100 \ \mu m \end{cases} \quad (6)$$

The above distribution can be considered as two concentric vortex lenses with focal length $f_1, f_2$ and topological charge $l_1, l_2$.

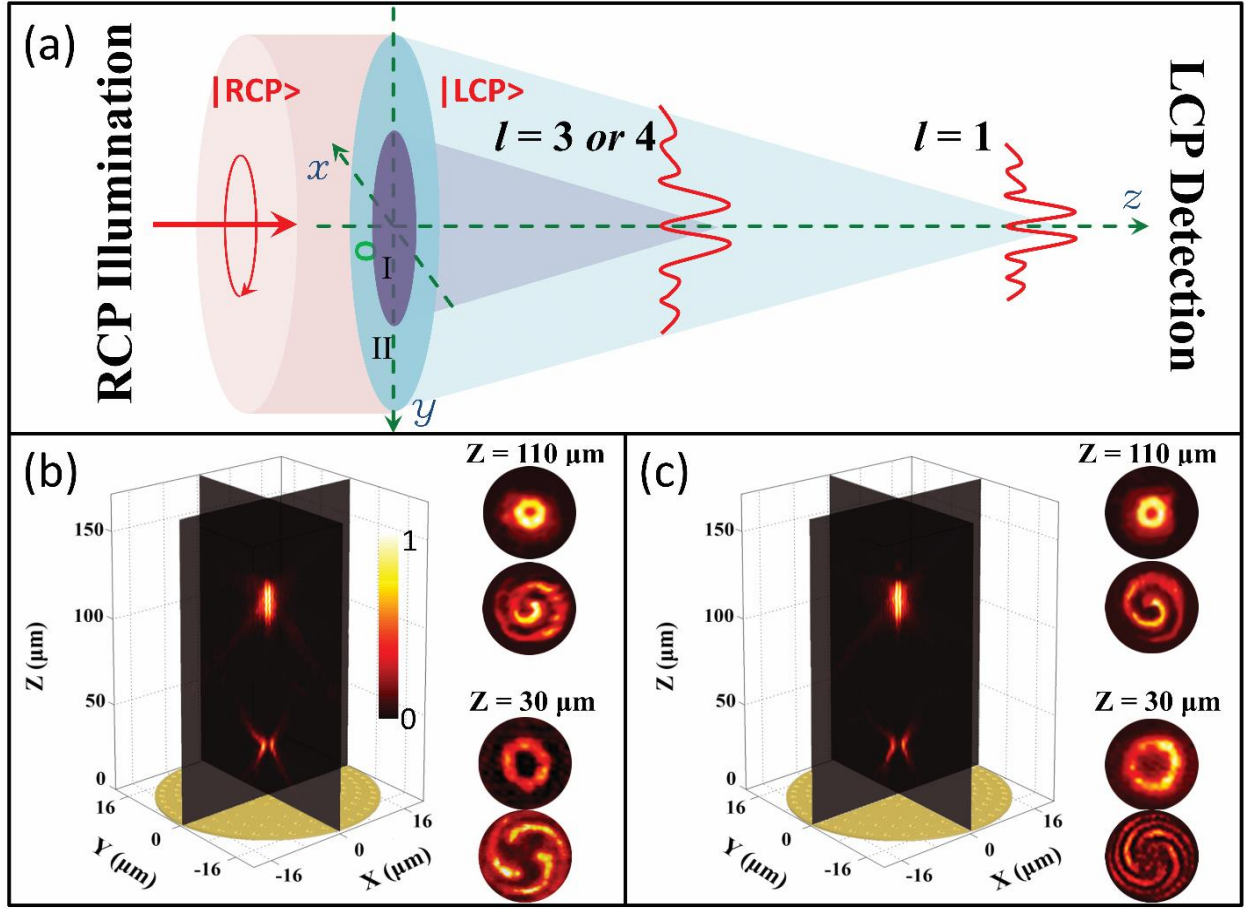

**Figure. 4.** Examples of multi-foci vortex lens (spatially multiplexing focused vortices). (a) Mechanism of the two-region vortex lens. (b) (c) Measured far-field longitudinal intensity distributions for vortex lens containing two regions encoded with different focal length ($f_1 = 30$ μm, and $f_2 = 110\ \mu m$) and different topological charge ($l_1 \neq l_2$). (b) $l_1 = 3$, $l_2 = 1$. (c) $l_1 = 4$, $l_2 = 1$. The right-side images in (b) and (c) show the intensity profiles and interference patterns of the beam's x-y cross-sections at the two different focal planes ($z = 30$ μm and $z = 110$ μm).

The two vortex lens designs are shown as following: (1) $l_1=3$, $f_1=30$ μm, $NA_1=0.83$, $l_2=1$, $f_2=110$ μm, $NA_2=0.67$; (2) $l_1=4$, $f_1=30$ μm, $NA_1=0.83$, $l_2=1$, $f_2=110$ μm, $NA_2=0.67$. In both designs, the two regions are designed for RCP illumination and LCP detection. Figure 4(a) shows the mechanism of the two-focus vortex lens, in which the two sub-vortex lenses focus two optical vortexes at two different focal planes ($f_1=30$ μm, $f_2=110$ μm). Figure 4(b) shows the measured data for the design of $l_1=3$, $l_2=1$, and Figure 4(c) shows the measured data for the design of $l_1=4$, $l_2=1$. At the second focal plane, it clearly shows quite similar patterns for the both designs because of the same parameters, while at the first focal plane, smaller topological charge [cf. Figure 4(b) $l_1$=3] leads to smaller-radius annular intensity pattern compared with the one carrying larger topological charge [cf. Figure 4(c) $l_1$=4] (refer to Figure 4S for the radial intensity distributions for both designs).

### 3.4. Dual-Polarity Vortex Lens

We also studied the polarization-dependent property of the two-region vortex lens. The two regions of the vortex lens are designed to response to the opposite circularly polarizations (dual-polarity vortex lens). The orientation angle distribution of the nano-voids can be rewritten as following:

$$\varphi(r) = \begin{cases} \pm\frac{\pi}{\lambda}\left(\sqrt{r^2 + f_1^2} - f_1\right) \pm \frac{l_1}{2}\tan^{-1}\left(\frac{y}{x}\right) & 0.5 \leq r \leq 45 \ \mu m \\ \mp\frac{\pi}{\lambda}\left(\sqrt{r^2 + f_2^2} - f_2\right) \pm \frac{l_2}{2}\tan^{-1}\left(\frac{y}{x}\right) & 50 \leq r \leq 100 \ \mu m \end{cases} \quad (7)$$

The only difference is the sign of the outside range shown in (7). More specifically, for the design of $l_1$=3, $l_2$=1 (Design 1), the inner sub-vortex lens is designed for RCP illumination, while the outer one is designed for LCP illumination. On the contrary, for the design of $l_1$=4, $l_2$=1 (Design 2), the inner sub-vortex lens is designed for LCP illumination, while the outer sub-vortex lens is designed for RCP illumination. Figure 5 shows the measured results of the dual-polarity vortex lenses. Under the illumination of RCP, in Design 1, the optical vortex with $l_1$=3 is focused at z = 30 μm (real image), and the optical vortex with $l_2$=1 is focused at z = -100 μm (virtual image) [cf. Figure 5(a)]. Oppositely, in Design 2, the optical vortex with $l_1$=4 is focused at z = -30 μm (virtual image), and the optical vortex with $l_2$=1 is focused at z = 100 μm (real image) [cf. Figure 5(b)]. Changing the polarization state of the incident light will flip over the focal planes, namely the real focal plane becomes virtual focal plane, and vice versa. This dual-polarity feature, enabling light with two opposite spins to possess the inverse divergence and resulting in the real and virtual focusing vortexes, may provide great convenience for nano-manipulation of optical vortex.

Benefitting from the Babinet design, the SNR for our nanosieve design is calculated to be -4 dB at working wavelength of 632.8 nm. All of the above mentioned designs can achieve about 3% efficiency with respect to the input power. It should be noted that the efficiency of the nanosieves can be easily enhanced by employing reflection-type versions [54] or dielectric versions [44, 55].

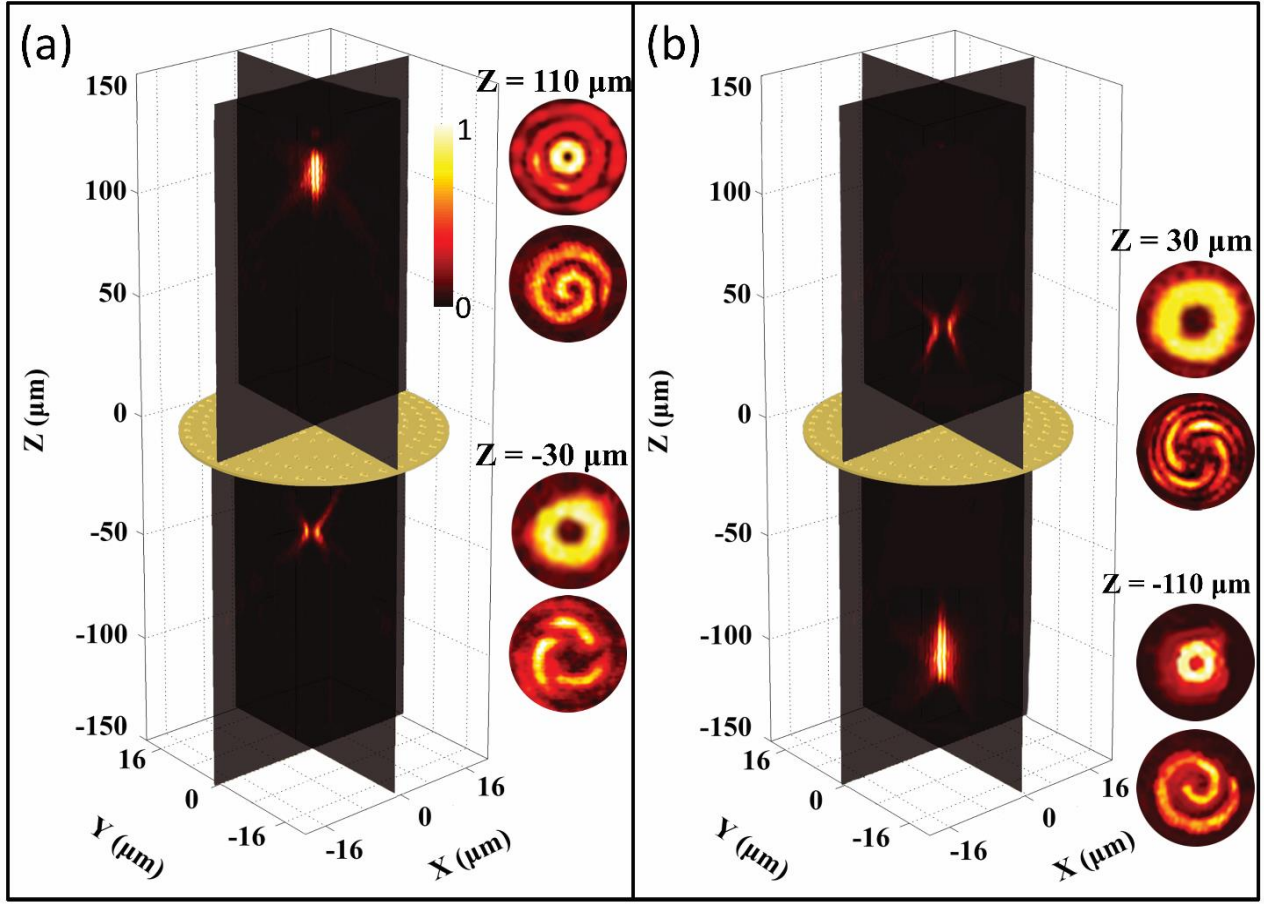

**Figure. 5.** Examples of dual-polarity vortex lens. Measured far-field longitudinal intensity distribution for vortex lens containing two regions with different focal length ($|f_1| = 30$ μm, and $|f_2| = 110\ \mu m$) and different topological charge ($l_1 \neq l_2$), and opposite polarities. (a) $l_1 = 3$ for RCP, $l_2 = 1$ for LCP. (b) $l_1 = 4$ for LCP, $l_2 = 1$ for RCP. The right-side images in (a) and (b) show the intensity profiles and interference patterns of the beam's x-y cross-sections at the two focal planes (|z| = 30 μm and |z| = 110 μm).

## 4. Conclusions

To summarize, our flat helical nanosieves device exhibits obvious advantages over traditional strategies on the aspect of complex manipulations of OAM, which possesses the merits of thinness, extensibility, and flexibility. In practice, although the highest topological charge is limited by the phase sampling frequency ($|l| < \frac{N}{2}$, $N$ is the average total number of antennas along the azimuthal direction), namely the number of the unit nano-sieve antenna, it can already meet the needs of majority of cases. It still has space to further decrease the pixel size to meet the needs in real cases, to say the least. On the aspect of an individual functional device, all of the mentioned benefits are shared, including the always desired high-purity generated field in virtue of polarization isolation. On the other hand, the strategy of merging several single-function devices into a single flat device deserves particular attentions, especially in the integrated systems. In addition, owing to the easy implementation of phase control, the situation of complex surface polishing, during combination of several functional phase response or realization of complex phase distribution (e.g. hologram retrieved phase) in traditional methods, is avoided.

To conclude, we have demonstrated the control of multiple vortex phase loaded light via designer flat helical nanosieves. This flat strategy can be extended onto many other multifunctional devices, for example, OAM readable holograms, which is another potential application during our following works.

## Experimental Section:

Au film with 60 nm thickness was deposited onto a quartz substrate by electron-beam vapor deposition (Denton Vacuum, Explorer) under a vacuum ($5 \times 10^{-7}$ Torr). Then positive resist (PMMA 950K A11) was coated, having mixed with anisole (1:3), via spin-coater (6000 rpm 45s). The coated sample was then baked at the hotplate for 2 min at 180 ℃. Electron beam lithography (eLine Plus, Raith) pattern the structure onto the resist by exposing the specimen at a dosage (280 μC cm$^{-2}$), beam current (47 pA) and with an acceleration voltage (20 kV). After patterning, the samples were developed for 40s in MIBK: IPA = 1:3. The specimen was then subjected to an argon ion beam etching process (Nanoquest, Intlvac) under conditions of acceleration voltage (45 V), beam voltage (300 V), beam current (100 mA), and RF power (170 W) for about 2 mins.

## Author Contributions.

Shengtao Mei and Muhammad Q. Mehmood contributed equally to this work. S.T.M., M.Q.M., and C.W.Q. conceived the idea. S.T.M. and M.Q.M. optimized the design through MATLAB and FDTD simulations. S.T.M, M.Q.M., and S.Y.S did all the fabrications. Characterizations and analyses of the measured results are done by S.T.M. and M.Q.M. S.T.M. and M.Q.M. prepared the manuscript, while K. H., X.L., H. L, A. D., C.-W.Q. and J.H.T. were involved in technical discussions and added fruitful suggestions/comments about the work. C.W.Q supervised the overall work in the project.

## Acknowledgment.

This research is supported by the National Research Foundation, Prime Minister's Office, Singapore under its Competitive Research Programme (CRP Award No. NRF-CRP10-2012-04). C.W.Q. also acknowledges support under the grant R-263–000-A45–112 from the National University of Singapore. X. Ling acknowledges the financial support of the National Natural Science Foundation of China under the grant 11447010. The work is also partially supported by Agency for Science, Technology and Research (A*STAR) under Grant Numbers 0921540099 and 1021740172.

# Flat Helical Nanosieves
# --Supplementary Materials

*Shengtao Mei[1, 3†], Muhammad. Qasim. Mehmood[1, 5†], Sajid Hussain[1†], Kun Huang[2], Xiaohui Ling[1, 4], Shawn.Yohanes.Siew[1], Hong Liu[2], Jinghua Teng[2], Aaron Danner[1], and Cheng-Wei Qiu[1, 3*]*

[1]Department of Electrical and Computer Engineering, National University of Singapore,
4 Engineering Drive 3, Singapore 117583, Singapore

[2]Institute of Materials Research and Engineering Agency for Science Technology and Research (A*STAR), #08-03, 2 Fusionopolis Way, Innovis, Singapore 138634

[3]Graduate School for Integrative Sciences and Engineering, National University of Singapore, Centre for Life Sciences (CeLS), #05-01, 28 Medical Drive Singapore 117456, Singapore

[4]College of Physics and Electronic Engineering, Hengyang Normal University, Hengyang 421002, China

[5]Information Technology University (ITU), 346-B, Ferozepur Rd, Lahore 54600, Pakistan.

[†]These authors contributed equally to this work

[*]Correspondence and requests for materials should be addressed to C.W. Q. (email: chengwei.qiu@nus.edu.sg).

## 1. Single-region helical axicons (measured results)

Single-region (Region I or Region II) helical axicons are first investigated to demonstrate the Bessel beam shaping performance of our helical nanosieves. To be more specific, a circular region (see Figure 1S (a) and 1S (b)) and an annular region (see Figure 1S (e) and 1S (f)) are investigated first (the flat helical axicons are placed at the plane of z = 0μm represent by black circular plane). The design parameters can be found in Table 1S.

The measured far-field nondiffracting beams for a circular flat helical axicon ($r_1 = 0.5\sim45$ μm) with $\beta = 6°, l = 1$ is shown in Figure 1S (a). The calculated DOF in this case is about $\frac{r_1}{\tan\beta} \approx 430$ μm which is well-verified by experiment result. Obviously, changing radius and opening angle both will influence the DOF. As shown in Figure 1S (b), the same radius circular flat helical axicon with $\beta = 3°, l = 1$ forms a much longer nondiffracting field along the optical path (DOF ≈ 860 μm). Considering annular flat helical axicons with $\beta = 6°, l = 3$, $r_1 = 0.5\sim45$ μm, and $r_2 = 50\sim100$ μm (Figure 1S (e)), and with $\beta = 6°, l = 4$, $r_1 = 0.5\sim45$ μm, and $r_2 = 50\sim100$ μm (Figure 1S (f)), both of the generated nondiffracting fields are from 470 μm to 950 μm, namely the verified value DOF ≈ 480 μm. Therefore, changing the topological charge *l* will not affect DOF. While the radius of the donut-shaped intensity is observed to be larger when the helical axicon is encoded with larger OAM, this is in consistent with the theoretical prediction. In addition, we observe the on-axis line of darkness in the right-side intensity figures that is the characteristic of the presence of an optical vortex. Meanwhile, the measured radial intensity profiles (circle markers) match well with theoretical predictions (solid curves: High order Bessel function fitting) for each design as shown in Figure 1S (c), (d), (g), and (h). Such highly-confined and nondiffracting vortex beams are not achievable by previous demonstrated plasmonic optical vortex.

**Table 1S** Parameters for designs in Figure 1S

|  | Inner sub-lens (Region I) | | Outer sub-lens (Region II) | |
|---|---|---|---|---|
|  | Fig 1S(a) | Fig 1S(b) | Fig 1S(c) | Fig 1S(d) |
| Opening Angle ($\beta$) | 6° | 3° | 6° | 6° |
| Number of Nano-voids | 20358 | | 101128 | |
| Radial Rotation Angle ($\varphi_A$) | [0.083$\pi$, 7.4$\pi$] | [0.041$\pi$, 3.7$\pi$] | [8.3$\pi$, 16.5$\pi$] | |
| Azimuthal Rotation Angle ($\varphi_S$) | [0, $\pi$] | | [0, 3$\pi$] | [0, 4$\pi$] |

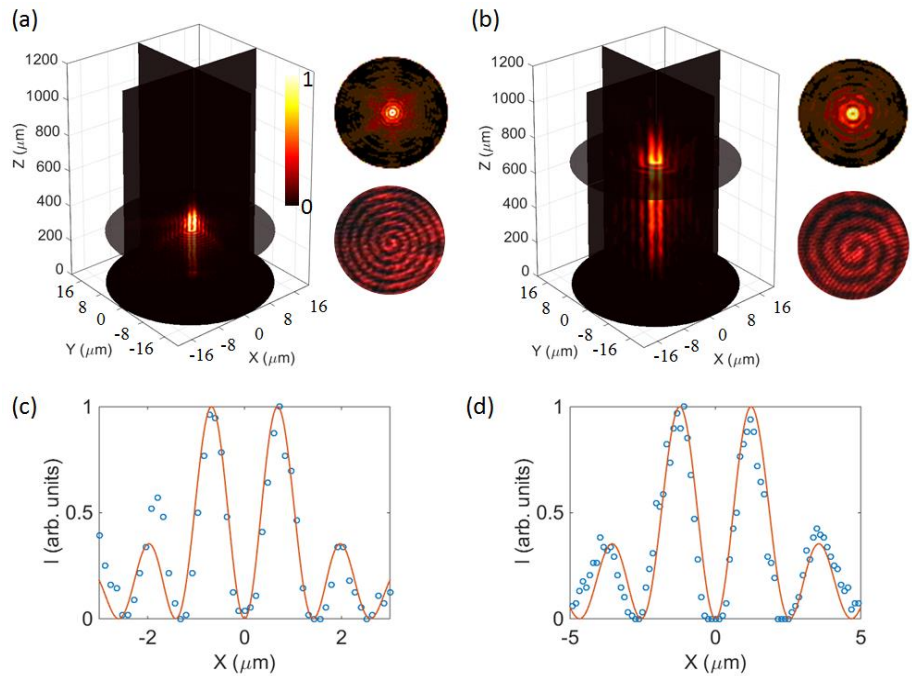

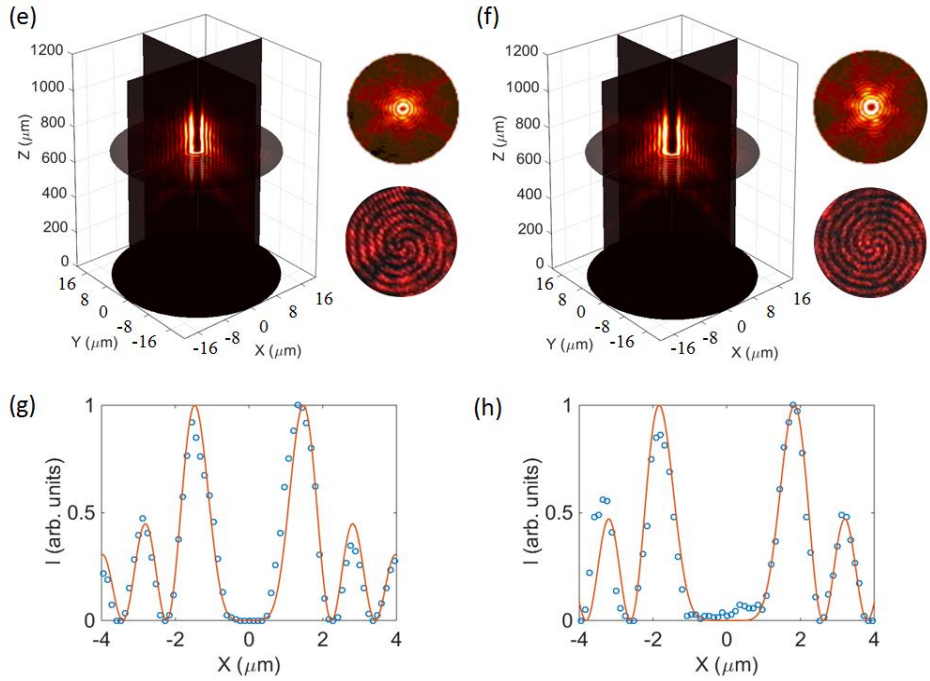

**Figure 1S** Measured far-field distribution along the propagating path for one-region helical axicons ((a), (b), (e), and (f)) and azimuthally averaged radial intensity profile ((c), (d), (f), and (g). The black circular planes lying at z = 0 μm represent the tested nanosieves devices (only part of the nanosieves device is shown). Simulation data: solid curves; Experiment data: circle markers). (a) $\beta = 6°, l = 1$. (b) $\beta = 3°, l = 1$. (e) $\beta = 6°, l = 3$. (f) $\beta = 6°, l = 4$; The right-side images in (a), (b), (e), and (f) show the intensity and interference pattern profiles of the beam's x-y cross sections (highlighted circular planes) within the nondiffracting field for each design (For (a), z = 300 μm; for others, z = 700 μm).

## 2. Double-region helical axicons (simulation results)

Two-region helicon axicons are also simulated to predict and support our measured results in the main text. The design parameters are shown in Table 2S and Table 3S. Figure 2S shows the helical axicon simulation results for the case of the two regions with the same opening angle ($\beta_1 = \beta_2 = 6°$). This simulation predicts the experimental results quite well. Figure 3S shows the simulation results for the case of two-region helical axicons with two different opening angles ($\beta_1 = 3°$, and $\beta_2 = 6°$), namely helicon wave generators. The rotating multi-lobe interference pattern result into the periodic segmented longitudinal field distribution, which is the pronounced behavior of helicon wave.

**Table 2S** Parameters for designs in Figure 2S

|  | Inner sub-lens (Region I) | | Outer sub-lens (Region II) | |
| --- | --- | --- | --- | --- |
|  | Fig 2S(a) | Fig 2S (b) | Fig 2S (a) | Fig 2S (b) |
| Opening Angle (β) | 6° | 6° | 6° | 6° |
| Number of Nano-voids | 20358 | | 101128 | |

| Radial Rotation Angle ($\varphi_A$) | [0.083π, 7.4π] | [8.3π, 16.5π] | |
|---|---|---|---|
| Azimuthal Rotation Angle ($\varphi_s$) | [0, π] | [0, 3π] | [0, 4π] |

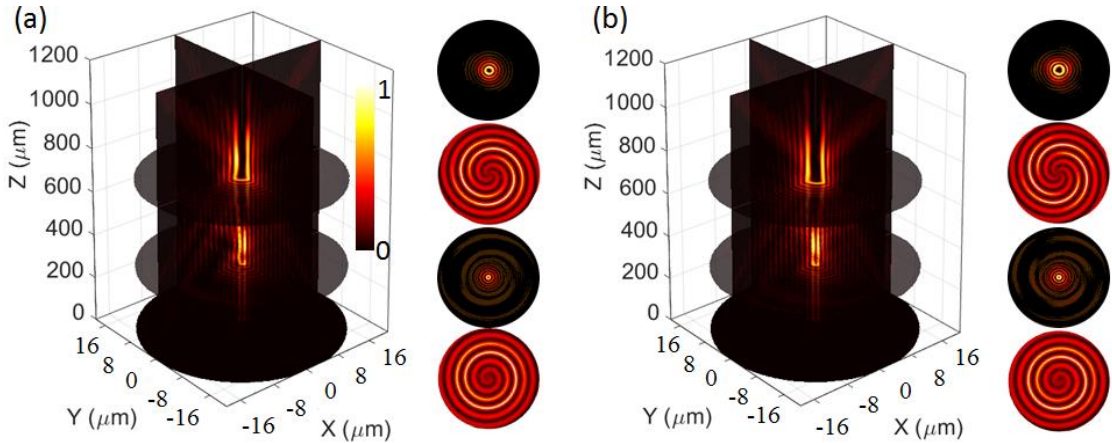

**Figure 2S** Simulations: (a) $l_1 = 1$, $l_2 = 3$. (b) $l_1 = 1$, $l_2 = 4$; The right-side images in (a) and (b) show the intensity profiles and interference patterns of the beam's x-y cross sections (highlighted circular planes) within the nondiffracting field for each design at z = 300 μm and z = 700 μm.

**Table 3S** Parameters for designs in Figure 3S

|  | Inner sub-lens (Region I) | | Outer sub-lens (Region II) | |
|---|---|---|---|---|
|  | Fig 3S(a) | Fig 3S(b) | Fig 3S(a) | Fig 3S(b) |
| Opening Angle (β) | 3° | 3° | 6° | 6° |
| Number of Nano-voids | 20358 | | 101128 | |
| Radial Rotation Angle ($\varphi_A$) | [0.08π, 3.7π] | | [8.3π, 16.5π] | |
| Azimuthal Rotation Angle ($\varphi_s$) | [0, π] | | [0, 3π] | [0, 4π] |

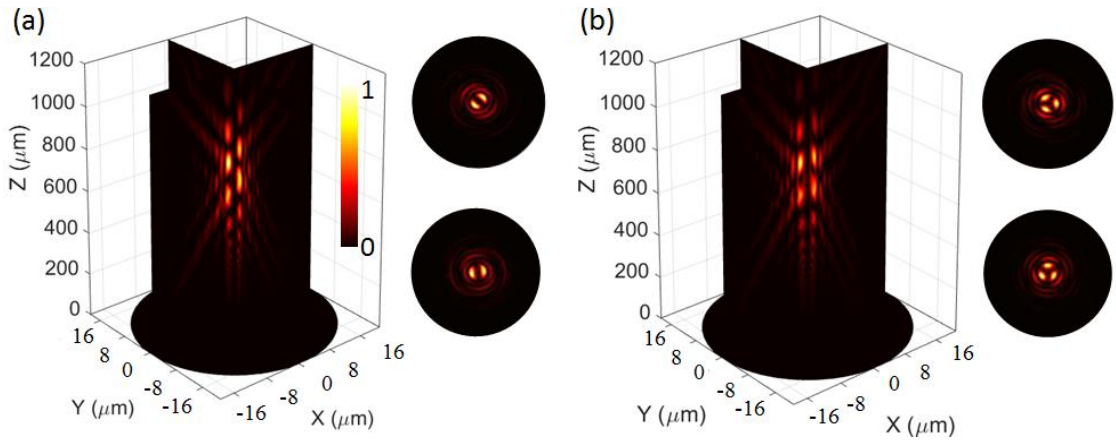

**Figure 3S** Simulations: (a) $\beta_1 = 3°, l_1 = 1,$ and $\beta_2 = 6°,\ l_2 = 3$. (b) $\beta_1 = 3°, l_1 = 1,$ and $\beta_2 = 6°,\ l_2 = 4$; The right-side images in (a) and (b) show the intensity profiles of the beam's x-y cross sections within the nondiffracting field for each design at z = 520 μm and z = 700 μm.

## 3. Single-region vortex lens (measured results)

We also experimentally investigate vortex lens with a single region (Region I or Region II). Figure 4S shows the measured single-region vortex lens's far-field longitudinal intensity distributions and radially averaged intensity distributions. To be more specific, a circular Region I (see Figure 4S (a)) and an annular Region II (see Figure 4S (b) and 4S (c)) are investigated first (The flat vortex lens is placed at the plane of z = 0μm). It can be seen that in (a) the radius of $l = 1$ is about 0.4 μm, in (b) the radius of $l = 3$ is about 1.1 μm, and in (c) the radius of $l = 4$ is about 1.3 μm. The larger the topological charge is, the bigger the radius of the annular focal intensity will be obviously. The specific parameters can be found in Table 4S

**Table 4S** Parameters for designs in Figure 4S

|  | Inner sub-lens (Region I) |  | Outer sub-lens (Region II) |
| --- | --- | --- | --- |
|  | Fig 4S(b) | Fig 4S(c) | Fig 4S(a) |
| Focal Length (*f*) | 30 μm |  | 110 μm |
| Number of Nano-voids | 20358 |  | 101128 |
| Radial Rotation Angle ($\varphi_L$) | [0.0056 π, 38π] |  | [44.7 π, 117.5π] |
| Azimuthal Rotation Angle ($\varphi_S$) | [0, 3π] | [0, 4π] | [0, π] |

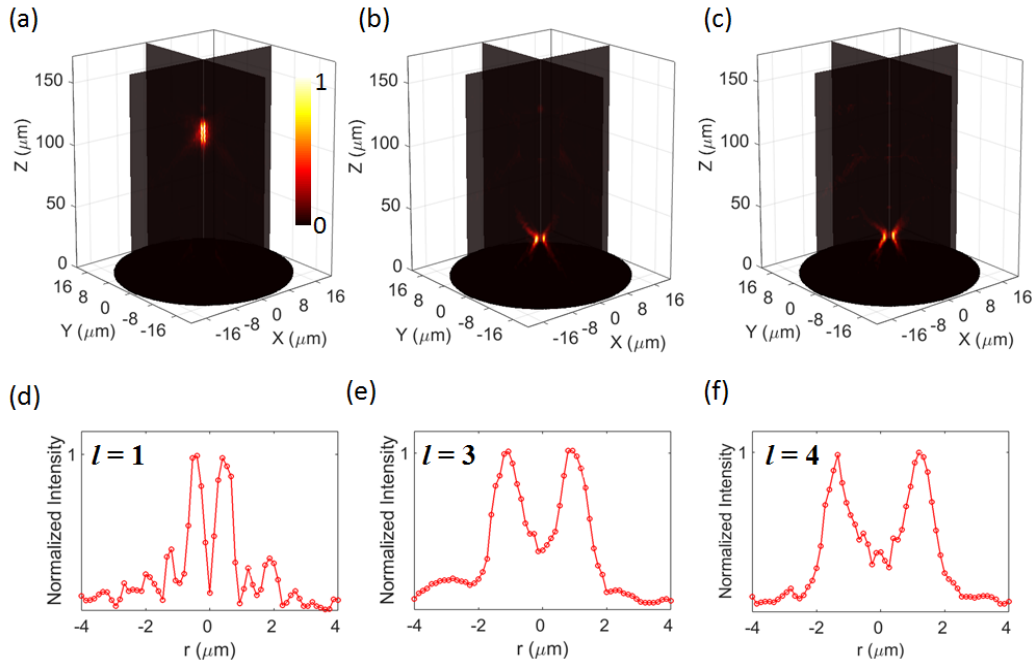

**Figure 4S** Measured single-region vortex lens's far-field longitudinal intensity distributions and radial intensity distributions at the focal plane for the fabricated three designs: (a) (d) $l = 1$, (b) (e) $l = 3$, (c) (h) $l = 4$.

## 4. Double-region vortex lens (simulation results)

Figure 5S shows the simulation example of a multi-foci vortex lens. There are two designs in this example, one with $l_1 = 3$, $l_2 = 1$ shown in Figure 5a and the other with $l_1 = 4$, $l_2 = 1$ shown in Figure 5b. Designing parameters can be found in Table 5S. It can be observed that even with higher NA ($NA_1 = 0.83 > NA_2 = 0.67$), OAM with larger topological charge

($l_1 = 3$ or $4$) cannot be focused smaller than the one with smaller topological charge ($l_2 = 1$), which reflects the effects of phase singularity.

Table 5S Parameters for designs in Figure 5S

|  | Inner sub-lens (Region I) | | Outer sub-lens (Region II) | |
| --- | --- | --- | --- | --- |
|  | Fig 5S(a) | Fig 5S(b) | Fig 5S(a) | Fig 5S(b) |
| Focal Length ($f$) | 30 μm | | 110 μm | |
| Number of Nano-voids | 20358 | | 101128 | |
| Radial Rotation Angle ($\varphi_L$) | [0.0056 π, 38π] | | [44.7 π, 117.5π] | |
| Azimuthal Rotation Angle ($\varphi_S$) | [0, 3π] | [0, 4π] | [0, 3π] | |

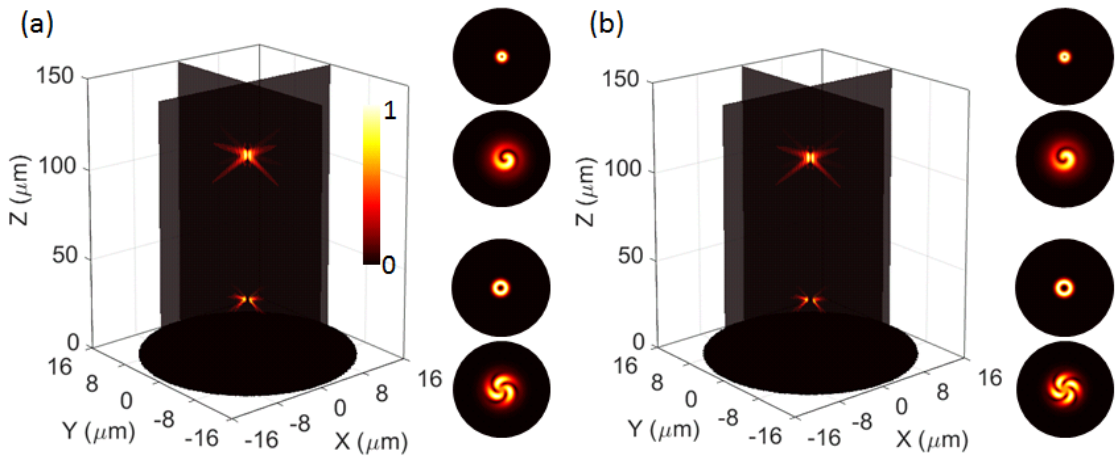

**Figure 5S** Example of a multi-foci vortex lens. (a) (b) Simulated far-field longitudinal intensity distributions for vortex lens containing two regions encoded with different focal length ($f_1 = 30\ \mu m$, and $f_2 = 110\ \mu m$) and different topological charge ($l_1 \neq l_2$). (a) $l_1 = 3$, $l_2 = 1$. (b) $l_1 = 4$, $l_2 = 1$. The right-side images in (a) and (b) show the intensity profiles and interference patterns of the beams' x-y cross-sections at the two focal planes ($z = 30\ \mu m$ and $z = 110\ \mu m$).

## 5. Characterization Method:

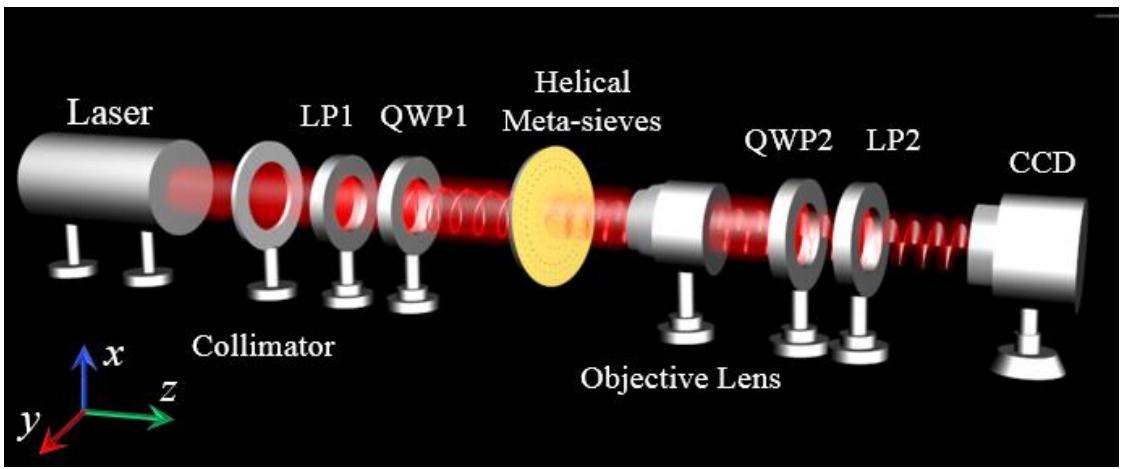

**Figure 6S** Experimental setup for optical characterizations. LP: linear polarizer, QWP: quarter waveplate

The experimental setup is shown in Figure 6S. He-Ne laser (λ = 633 nm) is used as the source, and it is collimated before LP. Two circular polarizers (linear polarizer and quarter waveplate) are used to transfer the incident light to be circularly polarized light and transfer the output light to be the opposite circularly polarized light, respectively. With the opposite circular polarizers, the designed beams can be observed at the CCD. If QWP2 or LP2 is tuned to let the background light (having the same charity with the incident light) transmit partially, a single-beam interferometric scheme can be achieved to test the interference pattern of the optical vortex beams.